\documentclass[12pt,a4paper]{article}

\usepackage{amsmath}
\usepackage{amsfonts}
\usepackage{amssymb}
\usepackage{amsthm}
\usepackage{slashed} 
\usepackage{graphicx}
\usepackage{epstopdf}
\usepackage[dvipsnames,svgnames,x11names,hyperref]{xcolor}
\usepackage[colorlinks=false]{hyperref}
\usepackage[figure,table]{hypcap} 
\hypersetup{
    bookmarksnumbered,
    pdfstartview={FitH},
  	citebordercolor={red},
    linkbordercolor={blue},
    urlbordercolor={ForestGreen},
    pdfpagemode={UseOutlines}
}

\newcommand{\ISpin}{\mathrm{ISpin}}
\newcommand{\Pin}{\mathrm{Pin}}
\newcommand{\Spin}{\mathrm{Spin}}

\newcommand{\SU}{\mathrm{SU}}
\newcommand{\U}{\mathrm{U}}
\newcommand{\OO}{\mathrm{O}}      
\newcommand{\SO}{\mathrm{SO}}
\newcommand{\ISO}{\mathrm{ISO}}

\newcommand{\dd}{\mathrm{d}}      
\newcommand{\D}{\mathrm{D}}

\newcommand{\R}{\mathbb{R}}

\newcommand{\vol}{\mathrm{vol}}

\newcommand{\tr}{\mathrm{tr}}
\newcommand{\diag}{\mathrm{diag}}
\newcommand{\Cl}{\mathrm{Cl}}
\newcommand{\bah}[1]{{\overline{#1}}}

\numberwithin{equation}{section}

\theoremstyle{definition}

\begin{document}

\title{Gauge theory of gravity and matter}

\author{Steven Kerr \\
Perimeter Institute for Theoretical Physics\\
31 Caroline Street North\\
Waterloo, Ontario, N2J 2Y5, Canada\\
\\
skerr@perimeterinstitute.ca
}

\date{}
\maketitle

\begin{abstract}
It is shown how to write the first order action for gravity in a gauge theoretic formalism where the spin connection and frame field degrees of freedom are assimilated together into a gauge connection. It is then shown how to couple the theory to spin-$0$, $\frac{1}{2}$, $1$ and $\frac{3}{2}$ fields in a gauge invariant fashion. The results hold in any number of spacetime dimensions.
\end{abstract}

\newpage

\section{Introduction}
Gauge theories have proven to be of great utility in describing fundamental physics. The standard model is a gauge theory of the group $\SU(3) \times \SU(2) \times \U(1)$. The first order formalism for gravity in $2+1$ dimensions is a Chern-Simons gauge theory of the group $\ISO(2,1)$ for zero cosmological constant and $\SO(3,1)/\SO(2,2)$ for positive/negative cosmological constant. In the language of gauge theory, the constraints of $(2+1)$-dimensional gravity become simple, and in \cite{Witten}, Witten exploited this fact in quantising the theory.

These developments and others have sparked interest in whether $(3+1)$-dimensional gravity can be described within the framework of gauge theory. In \cite{Kibble, Sciama}, Sciama and Kibble discovered that gravity in $3+1$ dimensions can be written in a form that exhibits a local $\SO(3,1)$ gauge symmetry using the frame field formalism. In the Macdowell-Mansouri formulation \cite{MM}, the action has an $\SO(4,1)/\SO(3,2)$ symmetry that is explicitly broken to obtain the Sciama-Kibble first order form of the gravitational action plus a topological term. In this approach, the frame field and spin connection are assimilated together as components of the $\SO(4,1)/\SO(3,2)$ connection. Stelle and West \cite{SW} expanded on this work by including an additional scalar field to carry out the symmetry breaking, thus restoring the overall $\SO(4,1)/\SO(3,2)$ symmetry. This action was improved by Pagels \cite{Pagels}, giving a simpler action principle and removing the topological term. Pagels also described how to couple the theory to scalar, spin-$\frac{1}{2}$ fermion, and Yang-Mills fields. A Poincar\'e group analogue of Pagels' work was developed by Grignani and Nardelli \cite{GN}, and the coupling to scalar and Yang-Mills fields improved upon by Ha \cite{Ha}.

In this paper, I review many of these developments with some additions. Here the $\ISO(n)$ invariant theories are written using a simplified matrix formalism, where $n$ denotes the number of spacetime dimensions. It is noted that it is possible to include a Holst type term \cite{Holst} within the gauge gravity formalism. The coupling to scalar and Yang-Mills fields proposed by Ha \cite{Ha} is generalised to the $\ISO(n)$ case. The spin-$\frac{1}{2}$ $\ISO(n)$ coupling proposed here differs from that of Grignani and Nardelli \cite{GN} in that the fermions transform trivially under translations and covariantly under rotations. The resulting action is in many ways simpler. The formalism is extended to include Rarita-Schwinger fields in arbitrary dimensions, and $N=1$ supergravity in $3+1$ dimensions. It is noted that it does not seem possible to obtain chiral fermions within the $\SO(n+1)$ theory.  All results generalise immediately to other group signatures, i.e. $\ISO(n-1,1)$, $\SO(n,1)$, etc.

It is shown that within this formalism, it is possible to describe fields of spin-$0$, $\frac{1}{2}$, $1$, $\frac{3}{2}$ and $2$ in any number of spacetime dimensions as a simple gauge theory.

\section{Gauge gravity action} \label{sec: Gauge gravity action}
In this section, the gauge action for gravity in $n$ spacetime dimensions is presented. For simplicity, we work with the $\SO(n+1)$ and $\ISO(n)$ theories. The $\SO(n+1)$ theory here was first reported by Pagels \cite{Pagels}. The $\ISO(n)$ action here is identical to that of Grignani and Nardelli \cite{GN}; however, here we use a matrix formalism that is similar to the $\SO(n+1)$ theory. The $\SO(n+1)$ theory naturally leads to the first order form of the Einstein-Hilbert action with a negative cosmological constant, while the $\ISO(n)$ theory naturally has zero cosmological constant. The signature of the groups may be altered to accommodate different spacetime signatures and a different sign of the cosmological constant, i.e. $\SO(n,1)$ or $\SO(n-1,2)$ as opposed to $\SO(n+1)$.

\subsection{$\SO(n+1)$ action}\label{sec:so(n) gauge gravity}

The Lie algebra of the group $\SO(n+1)$ can be represented as real, antisymmetric $(n+1) \times (n+1)$ matrices. The $\SO(n+1)$ connection is denoted $A^{BC}$, and the curvature is ${F}^{BC}=\dd A^{BC} +A^{BD}\wedge A_D^{\phantom{D}C}$, with upper case indices $A,B \ldots=0,\ldots,n$ and repeated indices contracted.
Indices are raised and lowered with the Euclidean metric $\delta_{AB}$. Let $\phi^A$ be a multiplet of scalar fields taking values in a sphere in $\R^{n+1}$ with constant radius $c>0$, so that $\phi^A\phi_A=c^2$. The covariant derivative is

\begin{align}
(\D\phi)^B=\dd \phi^B+{A^B}_C \phi^C. \label{covariant}
\end{align}
Pagels' action is \cite{Pagels}

\begin{align}\label{pagelsaction}
S_2 = \int & (\D \phi)^A\wedge (\D \phi)^B\wedge\ldots 
(\D\phi)^W \wedge  {F}^{XY} 
\epsilon_{AB\ldots WXYZ} \phi^Z. 
\end{align}
In $n$ spacetime dimensions, there are $(n-2)$ instances of $\D\phi$ in the action. For example, the $(3+1)$-dimensional action is 
 
\begin{align}
 \int & (\D \phi)^A 
\wedge  (\D \phi)^B 
\wedge {F}^{CD} 
\epsilon_{ABCDE} \phi^E. 
 \label{new action}
\end{align}

The $\phi$ field is a vector in $\R^{n+1}$ and so, by a gauge transformation, it may be rotated so that it points along the final coordinate axis,
\begin{align}\label{physicalgauge}
 \phi^Z \rightarrow \left( \begin{array}{c} 0 \\
 \vdots \\
 0 \\
 c \\
 \end{array} \right).
 \end{align}
This gauge choice is known as `physical gauge'. In this gauge, the fields may be written in terms of $\SO(n)$ tensors in block form, 

\begin{align}\label{physicalgauge2}
 \phi^Z=\begin{pmatrix}0\\c\end{pmatrix}
 \end{align}
with $0\in\R^n$, $c\in\R$, and

\begin{align}\label{connection}
A^{BC} =\begin{pmatrix}\omega^{bc}&e^b\\-e^c&0\end{pmatrix}. 
\end{align}
Here $\omega^{bc}$ is an $n\times n$ matrix of one-forms and $e^b$ is an $n$-dimensional vector of one-forms. Capital indices $A,B \ldots=0\ldots n$ are in the fundamental representation of $\SO(n+1)$, and the corresponding lower case indices $a,b \ldots =0\ldots (n-1)$ are in the fundamental representation of the $\SO(n)$ subgroup.  Thus $A^{bc}=\omega^{bc}$, $A^{bn}=e^b$, $A^{nc}=-e^c$ $A^{nn}=0$. 
 
In the physical gauge, the 1-forms $e^b$ and $\omega^{bc}$ are interpreted as the components of the frame fields and spin connection respectively. We have

\begin{align}
 (\D\phi)^B = \begin{pmatrix}ce^b\\ 0\end{pmatrix}.
 \end{align}
 Defining $R^{ab}$ to be the curvature of the  $\SO(n)$-connection $\omega$, we have
\begin{equation} F^{ab}=R^{ab}- e^a\wedge e^b.\end{equation}
Thus the action \eqref{pagelsaction} is
 \begin{align}\label{gravityaction}
 S_2 = c^{n-1} \int \left( e^a  \wedge e^b\wedge\ldots \wedge e^w \wedge R^{xy}\, \epsilon_{ab\ldots wxy} - e^a \wedge e^b \wedge\ldots\wedge e^y\, \epsilon_{ab\ldots y} \right) .
 \end{align}
This is the Sciama-Kibble action for gravity with a non-zero cosmological constant. It can be made to take its more familiar form by rescaling the frame field. Defining $\Lambda>0$ by $c^{n-1}\Lambda^{(n-2)/2}=1/G\hbar$ and setting $\tilde e^a=\Lambda^{-1/2}\,e^a$ results in the usual first-order form of the Einstein-Hilbert action with negative cosmological constant,
\begin{align}
 S_2 = \frac1{G\hbar} \int \left( \tilde e^a  \wedge \tilde e^b\wedge\ldots \wedge \tilde e^w \wedge R^{xy}\, \epsilon_{ab\ldots wxy} -\Lambda \,\tilde e^a \wedge \tilde e^b \wedge\ldots\wedge \tilde e^y\, \epsilon_{ab\ldots y} \right) .
 \end{align}
 The group signature may be altered to obtain the same action with a positive cosmological constant.

\subsection{$\ISO(n)$  action}\label{subsection: gauge gravity}

In this section, the gauge gravity action for the Euclidean group is constructed by analogy with the action of Pagels. This uses a matrix representation of $\ISO(n)$ that is similar to the defining representation of $\SO(n+1)$.  This results in an action in which the cosmological constant is naturally zero. When written in field components, the action coincides with the action studied in \cite{GN}. First some facts about the representation theory of the Euclidean group are reviewed.

\subsubsection{Representations of the Euclidean group}
The $n$-dimensional Euclidean group $\ISO(n)$ is the group of rotations and translations of $\R^n$. Its action is given by $x \rightarrow Mx + t$, where $M\in\SO(n)$ is the rotation matrix and $t\in\R^n$ is the translation vector. This defining representation has a non-linear action, but the group possesses a linear representation in $n+1$ dimensions given by
\begin{align}
\left( \begin{array}{c} x \\
c \\
\end{array} \right) &\rightarrow \left( \begin{array}{cc} M & t \\
0 & 1 \\
\end{array} \right) \left( \begin{array}{c} x \\
c \\
\end{array} \right) = \left( \begin{array}{c} Mx+ct \\
c \\
\end{array} \right). \label{vector rep}
\end{align}
This will be called the vector representation, and indices transforming in this representation will be denoted as upper indices. The particular value $c=0$ gives a linear subspace that is a sub-representation in which only the rotations act,
\begin{align}\label{quotientrep}
\begin{pmatrix}x\\0\end{pmatrix}\rightarrow\begin{pmatrix} Mx\\0\end{pmatrix}.
\end{align}

The dual of the vector representation is called the covector representation, and indices transforming in this representation will be denoted as lower indices. 
This may be represented as follows. Let $k=(k_0,\ldots,k_{n-1})\in\R^n$ and $k_n=\Omega\in\R$ be the last coordinate. The action is
\begin{align}
\left( \begin{array}{c} k \\
\Omega \\
\end{array} \right) &\rightarrow \left( \begin{array}{cc} M & 0 \\
-M^{-1}t & 1 \\
\end{array} \right) \left( \begin{array}{c} k \\
\Omega \\
\end{array} \right)= 
 \left( \begin{array}{c} Mk \\
-\left(M^{-1}t\right).k + \Omega \\
\end{array} \right). \label{covector rep}
\end{align}
The invariant contraction of a vector and covector is
\begin{align} x^Ak_A=x\cdot k + c\Omega=
 \left( \begin{array}{cc} Mx+ct & c  \\
\end{array} \right) \left( \begin{array}{c} Mk \\
-\left(M^{-1}t\right).k + \Omega \\
\end{array} \right) = x'^A k'_A .
\end{align}
The primes here denote transformed quantities. 

The invariant bilinear form that can be used to contract two covectors is given by
\begin{align}
\eta^{AB} = \left( \begin{array}{cccc} 1 & & &  \\
&  \ddots & & \\
& & 1& \\
& & & 0 \end{array} \right), \label{upper metric}
\end{align}
since $ k'_A \eta^{AB} l'_{A} = k'\cdot l'=k\cdot l = k_{A} \eta^{AB} l_A $.

The invariant bilinear form that can be used to contract two vectors is given by
\begin{align}
\eta_{AB} = \left( \begin{array}{cccc} 0 & & &  \\
 & \ddots & & \\
& & 0 & \\
& & & 1  \end{array} \right) , \label{lower metric}
\end{align}
since $ x'^A \eta_{AB} y'^A = c^2 = x^A \eta_{AB} y^A $.

Vectors that lie in the sub-representation \eqref{quotientrep} may be contracted using the identity matrix $\delta_{AB}$, since these vectors transform trivially under translations.

Finally, we note that the permutation symbols $\epsilon_{AB\ldots YZ}$ and $\epsilon^{AB \ldots YZ}$ are both invariant, since the transformations $\eqref{vector rep}$ and $\eqref{covector rep}$ both have determinant $1$.

\subsubsection{$\ISO(n)$  action}
Now an $\ISO(n)$ invariant action will be constructed using these ingredients.  Since the bilinear forms \eqref{upper metric}, \eqref{lower metric} are degenerate, the operations of raising and lowering indices are not invertible and one has to be work out which quantities are naturally vectors or covectors. 

The $\phi$ field is now a multiplet of real scalar fields in the vector representation, with a fixed constant $c\in\R$ as the last component,
$\phi^A = \left( \begin{array}{c} \phi^a \\
c \\
\end{array} \right) $. The $\ISO(n)$ connection is given in block form as

\begin{align}{A^B}_C=\begin{pmatrix}{\omega^b}_c &e^b\\0&0\end{pmatrix}. \label{iso connection}
\end{align}
The covariant derivative of $\phi^B$ is
\begin{equation}(\D\phi)^B=\begin{pmatrix} \dd\phi^b +{\omega^b}_c\phi^c + ce^b\\0\end{pmatrix}.\end{equation}
Since the last component is zero, this lies in the sub-representation \eqref{quotientrep}, transforming covariantly under rotations and not at all under translations.

Now consider the Euclidean field strength tensor, ${F}=dA + A\wedge A$. In terms of matrix components it is given by
\begin{align}
{F}^B_{\;\;\;C} = \left( \begin{array}{cc} \dd{\omega^b}_c + {\omega^b}_d \wedge {\omega^d}_c & \dd e^b + {\omega^b}_d \wedge e^d  \\
0 & 0  \end{array} \right).
\end{align}
Raising the second index using the metric $\eta^{AB}$ in $\eqref{upper metric}$ gives
\begin{align}
{F}^{BC} = {F}^B_{\;\;\;D} \eta^{DC} = \left( \begin{array}{cc} \dd \omega^{bc} + {\omega^b}_d \wedge \omega^{dc}  & 0  \\
0 & 0  \end{array} \right).
\end{align}
This tensor is also invariant under translations.

Finally an $\ISO(n)$ invariant action can now be constructed using the same formula as Pagels' action \eqref{pagelsaction},

\begin{align}
S_2 = \int & (\D \phi)^A\wedge (\D \phi)^B\wedge\ldots 
(\D\phi)^W \wedge  {F}^{XY} 
\epsilon_{AB\ldots WXYZ} \phi^Z.
\end{align}

This action can be gauge fixed, as in \eqref{physicalgauge}, whereupon the components of the $\ISO(n)$ connection \eqref{iso connection} are identified with the spin connection and the frame field. It reduces to
  \begin{align} \label{gravityaction2}
 S_2 = c^{n-1} \int e^a  \wedge e^b\wedge\ldots\wedge e^w \wedge R^{xy}\, \epsilon_{ab\ldots wxy}  ,
 \end{align}
which is exactly the Sciama-Kibble action for gravity with zero cosmological constant. The action allows a rescaling of the frame field, which is equivalent to changing the value of $c$. 

An independent cosmological term may be added to the both the $\ISO(n)$ and $\SO(n+1)$ theories by starting with an additional term in the action proportional to
\begin{align}
 \int \vol',
\end{align}
where the $n$-form $\vol'$ is defined as

\begin{align}
\vol'= (\D\phi)^A \wedge  (\D\phi)^B \wedge\ldots\wedge (\D\phi)^Y  \epsilon_{AB\ldots YZ} \phi^Z. \label{volume form}
\end{align}
There are $n$ instances of $D\phi$ in this formula. In the physical gauge, it is readily seen that $ \vol' = c^{n+1} \vol=  c^{n+1} e^a \wedge e^b \wedge\ldots\wedge e^y\, \epsilon_{ab\ldots y}$, with $\vol$ the canonical volume form on the spacetime manifold.

Finally we note that in $n=4$ spacetime dimensions, it is possible to include a Holst type term \cite{Holst} in both the $\SO(5)$ and $\ISO(4)$ theories,

\begin{align}
S_H = \int  (\D\phi)^A \wedge  (\D\phi)^B \wedge F_{AB}.
\end{align}
In the physical gauge, this reduces to

\begin{align}
S_H = \int  e^a \wedge e^b \wedge F_{ab}.
\end{align}

\section{Coupling to matter}\label{chapter: Coupling to matter}
In this section, the coupling of the gauge gravity action to matter is explored. A way of coupling the $\SO(n+1)$ theory to spin-$\frac{1}{2}$ fermion fields has been explored before by Pagels \cite{Pagels}, and subsequently the $\ISO(n)$ case was explored by Grignani and Nardelli \cite{GN}. The general form of the coupling to scalar and Yang-Mills fields was worked out by Ha in \cite{Ha} in the $\SO(n+1)$ case. We generalise this to the $\ISO(n)$ case. 

The spin-$\frac{1}{2}$ $\SO(n+1)$ coupling presented in this section is identical to that of Pagels. The spin-$\frac{1}{2}$ $\ISO(n)$ coupling presented in this section is simpler than that of Grignani and Nardelli insofar as the spinors transform covariantly under rotations and trivially under translations, rather than covariantly under the whole Euclidean group, and the action is written using the simplified matrix formalism of the previous section.

All results immediately generalise to any group signature, i.e. $\ISO(n-1,1)$, $\SO(n,1)$, etc.

\subsection{Bosons}
The general action for a real singlet scalar field $\sigma$ in $n$ dimensions is

\begin{align}
S_0 =  \frac{1}{2} &\int \vol \; \left[ (\partial^{\mu} \sigma) (\partial^{\nu} \sigma) g_{\mu \nu}  - m^2 \sigma^2 \right], \label{general scalar}
\end{align}
where $g_{\mu \nu}$ is the spacetime metric and $\vol$ is the canonical volume form.

The $\SO(n+1)$ theory may be coupled to a real singlet scalar field $\sigma$ in a gauge invariant way as follows,

\begin{align}
S_0 = \frac{1}{2} &\int {\vol'} \; \left[ (\partial^{\mu} \sigma) (\partial^{\nu} \sigma) g_{\mu \nu}'   - m^2 \sigma^2 \right]. \label{scalar coupling}
\end{align}
Here $\vol'$ is defined in \eqref{volume form}, and $g_{\mu \nu}'$ is defined by 

\begin{align}
g_{\mu \nu}' &= (D_{\mu} \phi)^A (D_{\nu} \phi)^B \delta_{AB}. \label{gauge metric}
\end{align}
Upon going to the physical gauge,

\begin{align}
 \phi^A \rightarrow \left( \begin{array}{c} 0 \\
 c \\
 \end{array} \right),
 \end{align}
with $c=1$ taken for convenience, and using the fact that the metric and frame field are related by

\begin{align}
g_{\mu \nu} = e_{\mu}^a e^b_{\nu} \delta_{ab},
\end{align}
it is seen that $g_{\mu \nu}' = g_{\mu \nu}$. From \eqref{volume form} we also have that $\vol' = \vol$. Therefore the action \eqref{scalar coupling} is equal to \eqref{general scalar}, the general action for a scalar field coupled to gravity.

This coupling may be straightforwardly generalised to the case where the scalar field is complex or charged under some additional symmetry group. In the latter case, the partial derivatives in \eqref{scalar coupling} are replaced by the appropriate covariant derivatives.

A similar strategy may be used to couple Yang-Mills theory to the gauge gravity action. The general action for a Yang-Mills field in $n$ dimensions is

\begin{align}
S_{1} = \int \vol \; \tr (F^{\mu \nu} F^{\rho \lambda}) g_{\mu \rho} g_{\nu \lambda}, \label{general Yang-Mills}
\end{align}
where $F^{\mu \nu}$ is the curvature tensor for an appropriate external symmetry group, and $\tr$ is the Killing form on the relevant Lie algebra.
The gauge gravity coupling is given by 

\begin{align}
S_{1}  =  &\int \vol' \; \tr(F^{\mu \nu} F^{\rho \lambda}) g_{\mu \rho}' g_{\nu \lambda}',
\end{align}
which immediately reduces to \eqref{general Yang-Mills} in the physical gauge with $c=1$.

Identical constructions works for the gauge group $\ISO(n)$. In that case, equation \eqref{gauge metric} is manifestly $\ISO(n)$ invariant because the vector $(\D \phi)^A$ lies in the sub-representation \eqref{quotientrep}, for which the identity matrix $\delta_{AB}$ is an invariant bilinear form.

\subsection{Fermions}
In this section, it is shown how to couple the gauge gravity action to fermion fields. First some preliminaries on the Clifford algebra.

Let the spacetime metric signature be $(p,q)$, with $p+q=n$. The gamma matrices $\gamma^a$ form a representation of the the real Clifford algebra $\Cl(p,q)$,

\begin{align}
\{ \gamma^a, \gamma^b \} = 2\eta^{ab}.
\end{align}
Here $\eta^{ab} = \diag (\overbrace{1,1, \ldots 1}^\text{p times}, \overbrace{ -1, -1, \ldots -1}^\text{q times})$. We will use the following useful notation for the gamma matrices,

\begin{align}
\gamma^{ab} &= \gamma^{[a} \gamma^{b]} = \frac{1}{2} \left( \gamma^a \gamma^b - \gamma^b \gamma^a \right)  \\
\gamma^{abc} &= \gamma^{[a} \gamma^b \gamma^{c]} = \frac{1}{6} \big( \gamma^{a}\gamma^b\gamma^{c} - \gamma^{a}\gamma^c \gamma^{b} \nonumber \\
& \quad \quad \quad\quad\quad\;\;\;+ \gamma^{b}\gamma^c\gamma^{a} - \gamma^{b}\gamma^a\gamma^{c} \nonumber  \\
& \quad \quad \quad\quad\quad\;\;\;+ \gamma^{c}\gamma^a\gamma^{b} - \gamma^{c}\gamma^b\gamma^{a} \big).
\end{align}
The gamma matrices act on spinors by matrix multiplication. The spin connection is given by $\omega = \frac{1}{2} \omega_{ab} \gamma^{ab}$, with $\frac{1}{2}\gamma^{ab}$ the generators of the spin group $\Spin(p,q)$. The spinor space is a vector space equipped with an inner product $\langle \psi, \psi' \rangle = \overline{\psi} \psi' = \psi^{\dagger} \gamma \hspace{0.25mm} \psi' $ that is preserved by $\Spin(p,q)$. The matrix $\gamma$ is hermitian and satisfies

\begin{align}
{\gamma^{ab}}^{\dagger} \gamma = - \gamma \hspace{0.3mm}\gamma^{ab} \;\; \forall a,b.
\end{align}
In an even number of spacetime dimensions, there are two possible independent solutions for $\gamma$. The matrix $\gamma$ may be proportional to the product of the hermitian gamma matrices $\prod_{a=0}^{p-1} \gamma^a$, or it may be proportional to the product of the antihermitian gamma matrices $\prod_{a=p}^{n-1} \gamma^a$. These two solutions are related up to a factor by multiplication by $\gamma^n \sim \prod_{a=0}^{n-1} \gamma^a$. In an odd number of spacetime dimensions, the matrix $\gamma^n$ is proportional to the identity matrix, and therefore the two solutions are not independent.

As noted in \cite{Pin}, the pin groups $\Pin(p,q)$ and $\Pin(q,p)$, which double cover $\OO(p,q)$ and $\OO(q,p)$ respectively, are not in general isomorphic. In particular, parity transformations on spinors (pinors) are represented as different operators in the two groups. There is also a sign ambiguity in the parity operator. Finally, in an even number of spacetime dimensions, there are two different surjective homomorphisms that map $\Pin(p,q)$ onto $\OO(p,q)$. However, in general a parity transformation $P_i$ which inverts the $i$-th spatial axis is given by either $\pm \gamma_i \gamma_n$ or $\pm \gamma_i$.

In an even number of spacetime dimensions, the two solutions for the inner product matrix $\gamma$ result in fermionic actions which transform differently under parity - one with a plus sign and the other with a minus sign. Suitable linear combinations of these parity symmetric and parity antisymmetric actions can be taken to obtain a theory of chiral fermions. This is equivalent to using the chirality projection operator to project out the different chiralities.

The components of the spinors form a Grassmann algebra, which is an algebra with an anti-commutative composition law. Hermitian conjugation of spinors is defined so that the combination $\overline{\psi}\psi$ is real - in other words, one does not pick up an additional minus sign from the Grassmann anti-commutation law in reversing the order of the factors upon conjugation.

\subsubsection{Spin $\frac{1}{2}$}
The massless Dirac action in $n$ spacetime dimensions is given by 

\begin{align}
S_{\frac{1}{2}}= \frac{\alpha}{2} \int  \;  e^a \wedge & \; e^b \wedge \ldots  e^x \wedge \left[ \overline{\psi} \gamma^y (d + \omega)\psi - \left((d + \omega) \overline{\psi}\right) \gamma^y  \psi \right]  \epsilon_{ab\ldots xy}. \label{Dirac action}
\end{align}
It general it is necessary to split the action into two terms to ensure that it is real because the torsion tensor is not assumed to be zero. There are $(n-1)$ instances of the frame field $e$ in this action. The spinor $\psi$ is in a representation of the spin group $\Spin(p,q)$. The integrand in \eqref{Dirac action} can be pure real or imaginary depending on the number of spacetime dimensions, the spacetime metric signature and the choice of spinor inner product. Therefore the constant $\alpha$ is chosen to be proportional to $1$ or $i$ appropriately so that the action is real. 

For convenience, we will construct actions principles that are invariant under $\ISO(n)$/$\SO(n+1)$ that are equivalent to the action \eqref{Dirac action}. However, identical conclusions hold for any other group signature.

\paragraph{\underline{$\SO(n+1)$ coupling:}} \label{fermion so(n+1)}

Consider the action

\begin{align}
S_{\frac{1}{2}}=  \frac{\alpha}{2} \int  &(\D\phi)^A \wedge (\D\phi)^B \ldots \wedge (\D\phi)^X \wedge \left[ \overline{\psi} \gamma^Y \D \psi - (\D \overline{\psi}) \gamma^Y  \psi \right] \epsilon_{AB \ldots XYZ} \phi^Z \label{even fermion action}
\end{align}
in $n$ spacetime dimensions, with the indices $A,B \dots$ in a representation of $\SO(n+1)$. There are $(n-1)$ instances of $\D\phi$ in this action. The spinor $\psi$ is in a representation of $\Spin(n+1)$. The gamma matrices form a representation of the real Clifford algebra $\Cl(n+1,0)$,

\begin{align}
\{ \gamma^A, \gamma^B \} = 2\delta^{AB}.
\end{align}
The action of the covariant derivative on the spinor $\psi$ is given by

\begin{align}
\D \psi = (\dd + \frac{1}{2} A_{BC} \gamma^{BC}) \psi,
\end{align}
where $A_{BC}$ are the real components of the $\Spin(n+1)$ connection and $\frac{1}{2} \gamma^{BC} = \frac{1}{2}(\gamma^B \gamma^C - \gamma^C \gamma^B)$ are the generators of $\Spin(n+1)$. The spinor inner product is given by $\bah\psi \psi = \psi^{\dagger} \Gamma \psi$, with $\Gamma$ a hermitian matrix satisfying

\begin{align}
{\gamma^{AB}}^{\dagger} \Gamma = - \Gamma \gamma^{AB} \;\; \forall A,B.
\end{align}

For now we assume $n$ is even. In that case the gamma matrices are taken to be $\gamma^Y = \left( \begin{array}{c} 
\gamma^a \\
\gamma^n \end{array} \right)$. The final gamma matrix is defined so that it is proportional to the product of all the others, $\gamma^n \sim \gamma^0 \gamma^1 \ldots \gamma^{n-1}$, with a possible factor of $i$ to ensure that it is hermitian. 

For $n$ even, the representation spaces of $\Spin(n+1)$ and $\Spin(n)$ are isomorphic as vector spaces, and are also isomorphic as inner product spaces upon choosing $\gamma := \Gamma$. Therefore the spinor $\psi$ may equally well be regarded as a spinor of $\Spin(n)$. The action \eqref{even fermion action} may be evaluated in the physical gauge,
 
 \begin{align}
 \phi^A \rightarrow \left( \begin{array}{c} 0 \\
 c \\
 \end{array} \right), 
 \end{align}
whereupon the components of the connection $A_{BC}$ are identified with the spin connection $\omega_{bc}$ and frame field $e_b$. Upon setting $c=1$, the action \eqref{even fermion action} is exactly equal to the generalised Dirac action \eqref{Dirac action}. It is possible to obtain the Dirac action with opposite parity by using the group $\SO(n,1)$. However, it is not possible to obtain chiral fermions by taking a linear combination of the two.

In an odd number of spacetime dimensions, the representation spaces of $\Spin(n+1)$ and $\Spin(n)$ are in general not isomorphic. However, in the Weyl representation the spinor $\psi$ is the direct sum of Weyl components $\chi_1$, $\chi_2$, each of which is in a representation of $\Spin(n)$,

\begin{align}
 \psi = \left( \begin{array}{c} \chi_1 \\
 \chi_2 \\
 \end{array} \right).
\end{align} 
Therefore for $n$ odd, the matrices $\gamma^Y$ are taken to be in the Weyl representation,

\begin{align}
\gamma^Y = \left( \begin{array}{cc} 0 & \sigma^Y \\
 \sigma^Y & 0 \\
 \end{array} \right), \label{Weyl rep}
\end{align}
where $\sigma^Y = \left( \begin{array}{c} \sigma^y \\
 1 \\
 \end{array} \right)$, and $\sigma^y$ generate the Clifford algebra $\mathrm{Cl}(n,0)$,
 
\begin{align}
\{ \sigma^a, \sigma^b \} = 2\delta^{ab}.
\end{align}
The inner product matrix is taken to be

\begin{align}
\Gamma = \left( \begin{array}{cc} 0 & \gamma \\
 \gamma & 0 \\
 \end{array} \right).
\end{align}
Evaluating the action \eqref{even fermion action} in the physical gauge with $c=1$ gives the generalised Dirac action \eqref{Dirac action} for two species of fermions $\chi_1$, $\chi_2$. The gauge invariant constraint $\chi_2=0$ may be imposed to obtain just one fermion species.

In an even number of spacetime dimensions, it is possible to have massive fermions by adding a term proportional to

\begin{align}
\int \vol' \; m \overline{\psi} \psi \label{mass term}
\end{align}
to the action \eqref{even fermion action}. This reduces to a mass term in the physical gauge. In an odd number of spacetime dimensions, the fermions are necessarily massless because a mass term for $\psi$ in the Weyl representation would mix the two fermion species $\chi_1$, $\chi_2$.

\paragraph{\underline{$\ISO(n)$ coupling:}} \label{fermion iso(n)}

The group $\ISpin(n) = \Spin(n) \ltimes \R^n$ is the semidirect product of $\Spin(n)$ with the translation group $\R^n$. A representation of the Lie algebra of $\ISpin(n)$ is given by

\begin{align}
\frac{1}{2}\gamma^{ab} = \frac{1}{4} \left( \gamma^a \gamma^b - \gamma^b\gamma^a \right), \quad\quad  t^c = 0, \label{ISO rep}
 \end{align}
where $t^c$ are the generators of $\R^4$. In this representation, the translations act trivially. These generators obey the Lie algebra of $\ISO(n)$,

\begin{align}
[t^a, t^b] &= 0, \\
[ \gamma^{ab}, \gamma^{cd}] &=  \delta^{ad} \gamma^{bc} + \delta^{bc} \gamma^{ad} - \delta^{ac} \gamma^{bd} - \delta^{bd} \gamma^{ac}  , \\
[ \gamma^{ab}, t^c ] &=  \delta^{bc} t^a - \delta^{ac} t^b.
\end{align}

The gamma matrices $\gamma^Y$ are in the sub-representation \eqref{quotientrep},

 \begin{align}
 \gamma^Y = \left( \begin{array}{c} \gamma^y \\
 0 \\
 \end{array} \right),
 \end{align}
where the final $Y=n$ component is the zero matrix. The covariant derivative is

\begin{align}
\D &= \dd + A^B_{\phantom{B}C} \gamma_B^{\phantom{B} C} \nonumber \\
&= \dd + \omega^b_{\phantom{b}c} \gamma_b^{\phantom{b} c},
\end{align}
where the connection $A^B_{\phantom{B}C}$ is given by \eqref{iso connection}, and the generators are $\frac{1}{2} \gamma_B^{\phantom{B} C} = \frac{1}{4} ( \gamma_B\gamma^C - \gamma^C \gamma_B ) $.

Evaluating the action \eqref{even fermion action} in the physical gauge, 

\begin{align}
 \phi^A \rightarrow \left( \begin{array}{c} 0 \\
 c \\
 \end{array} \right),
 \end{align}
and setting $c=1$ gives the generalised Dirac action \eqref{Dirac action}. A mass term can be included by addition of a term of the form \eqref{mass term} to the action \eqref{even fermion action}. There is no qualitative difference between the even and odd-dimensional cases as there was in the $\SO(n+1)$ case.

There are other possibilities for writing an action that is gauge invariant under the group $\ISO(n)$ and that reduces to \eqref{Dirac action} in the physical gauge. In particular, it is possible to have spinors that transform non-trivially under translations, which is achieved in \cite{GN}.

\subsubsection{Spin $\frac{3}{2}$}

The massless Rarita-Schwinger action for spin $\frac{3}{2}$ fields in $n$ spacetime dimensions is given by

\begin{align}
S_{\frac{3}{2}} = \frac{\beta}{2} \int e^a \wedge e^b \wedge \ldots e^v \wedge \left[ \bah\psi \gamma^{wxy} \wedge \D\psi + (\D\bah\psi) \gamma^{wxy} \wedge \psi \right] \epsilon_{ab \ldots vwxy}. \label{RS action}
\end{align}
Here all the symbols retain the same meaning except for the field $\psi = \psi_{\mu} \dd x^{\mu}$, which is now a spinor valued $1$-form with the suppressed spinor index transforming under $\Spin(p,q)$, $p+q=n$. The covariant derivative of $\psi$ is

\begin{align}
D_{\mu} \psi_{\nu} = \partial_{\mu} \psi_{\nu} + \frac{1}{2} \omega^{ab}_{\mu} \gamma_{ab} \psi_{\nu} + \Gamma_{\mu \nu}^{\phantom{\mu \nu}\rho} \psi_{\rho}, \label{covariant derivative}
\end{align}
where $\Gamma_{\mu \nu}^{\phantom{\mu \nu}\rho}$ are the affine connection coefficients. This connection is not assumed to be torsion-free, and so $\Gamma_{\mu \nu}^{\phantom{\mu \nu}\rho} \neq \Gamma_{\nu \mu}^{\phantom{\mu \nu}\rho}$. Indeed, the contributions from the torsion-free part of the connection in the above action vanish due to the antisymmetrisation, and only the contributions from the contorsion tensor survive. In general it is necessary to split the action into two terms to ensure that it is real because the torsion tensor is not assumed to be zero. There are $(n-3)$ instances of the frame field $e$ in this action. The integrand in \eqref{RS action} can be pure real or imaginary depending on the number of spacetime dimensions, the spacetime metric signature and the choice of spinor inner product. Therefore the constant $\beta$ is chosen to be proportional to $1$ or $i$ appropriately so that the action is real. 

The Rarita-Schwinger action \eqref{RS action} on its own leads to inconsistencies in the quantum theory. If $\mu$ in $\psi_{\mu}$ is taken as a $4$-vector index as above, then \eqref{RS action} is Lorentz invariant but the $\mu = 0$ components are non-dynamical and lead to negative norm states upon quantisation. In the flat space limit, the action \eqref{RS action} has a symmetry $\psi_{\mu} \rightarrow \psi_{\mu} + \partial_{\mu} \lambda $ for arbitrary spinors $\lambda$ that removes these degrees of freedom. However, in curved space there is no such symmetry available, and the problematic states persist. One way to overcome this obstacle is to combine \eqref{RS action} with some other fields in a supersymmetric fashion, which restores the necessary gauge symmetry. In that case, $\psi_{\mu}$ does not transform as a $4$-vector but as a connection, and there should be no contribution from the affine connection in the covariant derivative \eqref{covariant derivative}. Thus the Rarita-Schwinger type actions in this subsection can, with minimal modification, be thought of as a single term in an action that is free of any consistency issues.

For convenience, we will construct actions principles that are invariant under $\ISO(n)$/$\SO(n+1)$ that are equivalent to the action \eqref{RS action}. However, identical conclusions hold for any other group signature.

\paragraph{\underline{$\SO(n+1)$ coupling:}} Consider the action

\begin{align}
S_{\frac{3}{2}} = \frac{\beta}{2} \int (\D \phi)^A \wedge (\D \phi)^B \wedge \ldots (\D \phi)^V \wedge \left[ \bah\psi \gamma^{WXY} \wedge \D\psi + (\D\bah\psi) \gamma^{WXY} \wedge \psi \right] \epsilon_{AB \ldots VWXYZ} \phi^Z. \label{RS gauge action}
\end{align}
in $n$ spacetime dimensions, with the indices $A,B \dots$ in a representation of $\SO(n+1)$. There are $(n-3)$ instances of $\D\phi$ in this action. The spinor index of $\psi$ is in a representation of $\Spin(n+1)$. The gamma matrices form a representation of the real Clifford algebra $\Cl(n+1,0)$,

\begin{align}
\{ \gamma^A, \gamma^B \} = 2\delta^{AB}.
\end{align}
The action of the covariant derivative on the spinor $\psi$ is given by

\begin{align}
D_{\mu} \psi_{\nu} = \partial_{\mu} \psi_{\nu} + \frac{1}{2} A_{\mu}^{BC} \gamma_{BC} \psi_{\nu} + \Gamma_{\mu \nu}^{\phantom{\mu \nu}\rho} \psi_{\rho},
\end{align}
where $A^{BC}$ are the real components of the $\Spin(n+1)$ connection and $\frac{1}{2} \gamma_{BC} = \frac{1}{2}(\gamma_B \gamma_C - \gamma_C \gamma_B)$ are the generators of $\Spin(n+1)$. The spinor inner product is given by $\bah\psi \psi = \psi^{\dagger} \Gamma \psi$, with $\Gamma$ a hermitian matrix satisfying

\begin{align}
{\gamma^{AB}}^{\dagger} \Gamma = - \Gamma \gamma^{AB} \;\; \forall A,B.
\end{align}

Upon making the relevant identifications as in ~\ref{fermion so(n+1)}, the action \eqref{RS gauge action} reduces to \eqref{RS action}. In an odd number of spacetime dimensions, the spinor index of $\psi_{\mu}$ is in the Weyl representation \eqref{Weyl rep} of $\Spin(n+1)$, and one naturally obtains two Rarita-Schwinger fields $\chi_{1\mu}$, $\chi_{2\mu}$. The gauge invariant constraint $\chi_{2\mu}=0$ can be imposed to obtain just one.

In an even number of spacetime dimensions, it is possible to have a massive Rarita-Schwinger field by adding a term proportional to

\begin{align}
\int (\D\phi)^A \wedge (\D\phi)^B \wedge \ldots (\D\phi)^W \wedge \bah\psi \gamma^{XY} \wedge \psi \epsilon_{AB\ldots WXYZ} \phi^Z \label{RS mass term}
\end{align}
to the action \eqref{RS gauge action}. This reduces to a Rarita-Schwinger mass term in the physical gauge,

\begin{align}
\int e^a \wedge e^b \wedge \ldots e^w \wedge \bah\psi \gamma^{xy} \wedge \psi \epsilon_{ab\ldots wxy}.
\end{align}
In odd dimensions, the fermions are necessarily massless because a mass term for $\psi_{\mu}$ in the Weyl representation would mix the two Rarita-Schwinger fields $\chi_{1\mu}$, $\chi_{2\mu}$.

\paragraph{\underline{$\ISO(n)$ coupling:}}

In the $\ISO(n)$ case, the spinor index of the Rarita-Schwinger field $\psi_{\mu}$ is taken to be in the representation \eqref{ISO rep} of $\ISpin(n)$, in which the translations act trivially.
The gamma matrices $\gamma^Y$ are in the sub-representation \eqref{quotientrep},

 \begin{align}
 \gamma^Y = \left( \begin{array}{c} \gamma^y \\
 0 \\
 \end{array} \right),
 \end{align}
where the final $Y=n$ component is the zero matrix. The covariant derivative is

\begin{align}
D_{\mu} \psi_{\nu} &= \partial_{\mu} \psi_{\nu} + A^B_{\mu C} \gamma_B^{\phantom{B} C} \psi_{\nu} + \Gamma_{\mu \nu}^{\phantom{\mu \nu}\rho} \psi_{\rho}  \nonumber \\
&= \partial_{\mu} \psi_{\nu} + \omega^b_{\mu c} \gamma_b^{\phantom{b} c} \psi_{\nu} + \Gamma_{\mu \nu}^{\phantom{\mu \nu}\rho} \psi_{\rho},
\end{align}
where the connection $A^B_{\phantom{B}C}$ is given by \eqref{iso connection}, and the generators are $\frac{1}{2} \gamma_B^{\phantom{B} C} = \frac{1}{4} ( \gamma_B\gamma^C - \gamma^C \gamma_B ) $.

Evaluating the action \eqref{RS gauge action} in the physical gauge, 

\begin{align}
 \phi^A \rightarrow \left( \begin{array}{c} 0 \\
 c \\
 \end{array} \right),
 \end{align}
and setting $c=1$ gives the generalised Rarita-Schwinger action \eqref{RS action}. A mass term can be included by addition of a term of the form \eqref{RS mass term} to the action \eqref{RS gauge action}. There is no qualitative difference between the even and odd-dimensional cases as there was in the $\SO(n+1)$ case.

\subsection{Supergravity}

The action for $N=1$ supergravity in $3+1$ dimensions can be obtained by adding together the $\ISO(4)$ actions for spin-$2$ and $\frac{3}{2}$ fields,

\begin{align}
S_{SG} = \frac{1}{4} S_2 + \frac{1}{2} S_{\frac{3}{2}}.
\end{align}
Here the actions $S_2$ and $S_{\frac{3}{2}}$ are understood to be the gauge actions \eqref{pagelsaction} and \eqref{RS gauge action} in $3+1$ dimensions. The fermions in the spin-$\frac{3}{2}$ action are Majorana, and the covariant derivative acting on the Rarita-Schwinger field does not contain an affine connection term. This action is supersymmetric \cite{SG}. In the $\SO(5)$ case, it is necessary to include a non-zero mass term for the Rarita-Schwinger field to maintain supersymmetry, since in that case the gauge gravity action naturally produces a non-zero cosmological constant \cite{SG cosmo}. It may be possible to include extended supersymmetries and supersymmetries in higher dimensions within this formalism, but this is not undertaken here.

\section{Conclusion}
In this paper, the Sciama-Kibble action for gravity in $n$ spacetime dimensions is written as a gauge theory of the group $\SO(n+1)/\ISO(n)$ for Euclidean metric signature, with the frame field degrees of freedom assimilated into the gauge connection. It is shown how to couple the theory to scalar, Yang-Mills, and spin-$\frac{1}{2}$ and spin-$\frac{3}{2}$ fermionic fields within the gauge-theoretic formalism. The gravitational actions explored here were first noted by Pagels \cite{Pagels} in the $\SO(n+1)$ case, and Grignani and Nardelli \cite{GN} in the $\ISO(n)$ case. The coupling of scalar and Yang-Mills fields to the $\SO(n+1)$ theory was first noted by Ha \cite{Ha}. The coupling of the $\SO(n+1)$ theory to spin-$\frac{1}{2}$ fermions was first noted by Pagels \cite{Pagels}. This paper has a number of novelties. The $\ISO(n)$ theories are written using a simplified matrix formalism. The coupling of spin-$\frac{1}{2}$ fermions to the $\ISO(n)$ theory is simpler than that proposed by Grignani and Nardelli \cite{GN} insofar as the spinors are placed in a representation of the gauge group that transforms trivially under translations. The coupling to scalar and Yang-Mills fields constructed by Ha \cite{Ha} in the $\SO(n+1)$ case is generalised to the $\ISO(n)$ case. The formalism is extended to include spin-$\frac{3}{2}$ Rarita-Schwinger fields, and $N=1$ supergravity in $3+1$ dimensions. It is noted that it is possible to introduce a Holst type term in the gravitational action. It is also noted that it does not appear possible to obtain chiral fermions within the $\SO(n+1)$ theory. For this reason, at present the $\ISO(n)$ formalism is tentatively favoured. All results here generalise immediately to other group signatures. 

Gauge theory has proven useful in quantising gravity in $2+1$ dimensions, and it is hoped that the current formalism might shed some light on the quantisation of gravity along with matter fields in higher dimensions.

\section{Acknowledgements}
I thank John Barrett for many useful discussions. This work was funded by the Leverhulme Trust.

\end{document}